\documentclass[prb]{revtex4} % style for Physical Review B and AJP are similar
\usepackage[dvips]{color,graphicx}
\usepackage{amsfonts}
\usepackage{amssymb}
\usepackage{latexsym}

\newcommand{\dalm}{\kern1pt\vbox{\hrule height 0.9pt\hbox{\vrule width
0.9pt\hskip 2.5pt\vbox{\vskip 5.5pt}\hskip 3pt\vrule width 0.3pt}\hrule height
0.3pt}\kern1pt}

\newcommand{\lw}[1]{\smash{\lower2.ex\hbox{#1}}}

\begin{document}

%\thispagestyle{empty}

%<<<<<<<<<<<<< TITLE >>>>>>>>>>>>>>>%
\title{A novel derivation of the rotating black hole metric}
%<<<<<<<<<<<<< AUTHOR >>>>>>>>>>>>>>>%
\author{Naresh Dadhich\thanks{Electronic address:nkd@iucaa.ernet.in}}
\email{nkd@iucaa.ernet.in}
%<<<<<<<<<<<<< ADDRESS >>>>>>>>>>>>>>>%
\affiliation{Centre for Theoretical Physics, Jamia Millia Islamia, New Delhi-110025, India}
\affiliation{Inter-University Centre for Astronomy \& Astrophysics, Post Bag 4, Pune~411~007, India}
\date{\today}

%======================================%
%<<<<<<<<<<<<< ABSTRACT >>>>>>>>>>>>>>>%
%======================================%
\begin{abstract}
We derive the rotating black hole metric by appealing to ellipsoidal symmetry of space and a general guiding principle of incorporation of the Newtonian acceleration for massive and no acceleration for massless particles.
\end{abstract}

%<<<<<<<<<<<<< PACS NUMBER >>>>>>>>>>>>>>>%
\pacs{04.20.-q, 04.20.Cv, 04.50.Gh, 11.10.-z, 11.15.-q}

\maketitle

\section{Introduction}

Any covering theory should always include the original theory in
the first approximation simply because the new theory arises by
enlarging the framework or by inclusion of some new property which was not included in the existing theory.
Einstein's theory of gravity includes additional properties of
self interaction of gravitational field as well as its linkage to zero
mass particles - light. It is natural to expect that it should approximate to
Newtonian gravity when these new features are not very significant.
It should be noted that inclusion of light in the gravitational interaction requires that gravity
must curve space. This is because velocity of light cannot change,
it can then feel gravity only if space is curved \cite{d0}. Thus gravity must curve space, the question is, which of
its property should do that? It should naturally be the additional property of self interaction - gravitational
interaction of gravitational field \cite{d,d1}. This means Newtonian law of gravity remains intact and it is
only that space is not flat but curved. Hence it is natural to expect Newtonian acceleration for massive particles
in the first approximation and light should simply freely float on the curved space \cite{d} following the natural geodetic motion. This
is the motivation and justification for the general guiding principle that massive particles should in the first
approximation feel Newtonian acceleration while massless ones should feel no acceleration. This is the principle that we wish to employ to obtain black hole metrics. \\

We had successfully applied this principle for obtaining the metric for a static black hole and we now
wish to apply it to the more involved case of a rotating black hole \cite{d}. This is the main object of
this note. However there also exists another insightful derivation of Kerr solution \cite{df}. In the next section we derive the Kerr metric for a rotating black hole which is followed by the
inclusion of electric charge on the hole giving rise to the Kerr-Newman metric. We end with a discussion. \\

\section{Kerr metric}

It is natural that the appropriate symmetry for the field of a static
black hole is spherical while it is ellipsoidal for the rotating black hole. That is  constant potential surfaces are spheres for
the former while they are ellipsoids for the latter. The Schwarzshild
metric for static black hole was derived \cite{d} by writing the metric
in spherically symmetric form and then demanding that radially falling
photon experiences no acceleration while the massive particle experiences
the Newtonian acceleration. This happens because the former requirement
determines $g_{tt}g_{rr} = -1$ while the latter fixes $g_{tt} = 1+2\Phi$
where $\Phi = -M/r$ is the Newtonian potential. We would like to employ
the same approach to obtain the Kerr metric for a rotating black hole. \\

It has been shown that the appropriate symmetry for a rotating source
is ellipsoidal \cite{kra} and it is given by the $3$-metric,

\begin{equation}
 dl^2 = \frac{\rho^2}{r^2+a^2} dr^2 + \rho^2 d\theta^2 + (r^2+a^2)\sin^2\theta d\phi^2
\end{equation}

where $\rho^2 = r^2 + a^2 \cos^2\theta$ and $a$ is a constant. It is
flat space where the metric assumes this form for ellipsoidal symmetry,
and the corresponding $4$-spacetime metric would be given by

\begin{equation}
 ds^2 = dt^2 -\frac{\rho^2}{r^2+a^2} dr^2 - \rho^2 d\theta^2 - (r^2+a^2)\sin^2\theta d\phi^2
\end{equation}

which could be transformed to the familiar Boyer-Lindquist coordinates

\begin{equation}
 ds^2 = A (dt - a \sin^2\theta d\phi)^2 - \frac{1}{A} dr^2 -\rho^2 d\theta^2 - \frac{\sin^2\theta}{\rho^2}[(r^2+a^2)d\phi - a dt]^2 \,
\end{equation}
where
\begin{equation}
A = \frac{r^2+a^2}{\rho^2}.
\end{equation}

This is exactly what we get when we set $M=0$ in the Kerr solution. Note that the
above form of the metric is completely free of gravitational dynamics
as it simply reflects the axial symmetry of the ellipsoidal geometry
of flat spacetime. We shall now proceed to inject it with gravitational field
through the mass of the hole. To describe the gravitational field of
rotating black hole what we need is to bring in gravitational potential
in $A$. In doing so, note that the angular part of the metric as well
as the angular dependence should remain intact as they are entirely dictated by the
ellipsoidal symmetry of space. We thus write

\begin{equation}
 ds^2 = \frac{f(r)}{\rho^2}(dt - a \sin^2\theta d\phi)^2 - \frac{\rho^2}{g(r)}dr^2 -\rho^2d\theta^2 - \frac{\sin^2\theta}{\rho^2}[(r^2+a^2)d\phi - a dt]^2.
\end{equation}

We have now to determine the functions $f(r)$ and $g(r)$ which would
bring in the gravitational dynamics. Since they are free of $\theta$, it would suffice to consider
motion along the axis to determine them. For timelike and null geodesics,
we write the energy constant of motion,

\begin{equation}
\frac{f}{\rho^2} \dot{t} = E.
\end{equation}
The Lagrangian for $\theta=0$,
\begin{equation}
{\cal L} = \mu^2 = \frac{f(r)}{\rho^2} \dot{t}^2 - \frac{\rho^2}{g(r)} \dot{r}^2
\end{equation}
which in view of Eq. (6) implies
\begin{equation}
\dot{r}^2 = (\frac{\rho^2}{f}E^2 - \mu^2)\frac{g}{\rho^2}
\end{equation}
where $\mu$ is mass of particle and $\dot{r}$ denotes derivative
w.r.t proper time and an affine parameter respectively for
$\mu\neq0$ and $\mu=0$. \\

Now demanding that photon experiences no acceleration;i.e. $\ddot{r}=0$
for $\mu=0$, we have
\begin{equation}
\ddot{r} = (\frac{g}{f})^{\prime}E^2 = 0
\end{equation}
where a prime denotes derivative relative to $r$. Clearly $g/f =
const. =1$, the constant is set equal to $1$ for asymptotic flatness, the
metric should go over to the Minkowski flat as $r\to\infty$. So we have
$f(r) = g(r)$ which we will determine from inclusion of the Newtonian
acceleration in timelike motion. The acceleration experienced by
a timelike particle with $\mu=1$ is given by
\begin{equation}
\ddot{r} = -\frac{(r^2+a^2){f}^{\prime} - 2rf}{2(r^2+a^2)^2}.
\end{equation}
Note that $f$ is a function of $r$ and it should reduce to $r^2 + a^2$
asymptotically so as to go over to flat spacetime, and hence we write
$f(r) = r^2+a^2 + \psi(r)$ and then the above equation takes the form,
\begin{equation}
\ddot{r} = -\frac{\psi^{\prime}}{2(r^2+a^2)} + \frac{r\psi}{(r^2+a^2)^2}
\end{equation}
which at large $r$ approximates to
\begin{equation}
\ddot{r} = -\frac{r\psi^{\prime} - 2\psi}{2r^3}.
\end{equation}
This should reduce to the Newtonian acceleration, $-M/r^2$ and hence
$\psi = -2Mr$. So the function $f(r)$ is determined to read as
\begin{equation}
f(r) = r^2 + a^2 -2Mr = g(r).
\end{equation}

With this put in Eq. (5), we have obtained the Kerr metric for a rotating black hole
of mass $M$ and specific rotation parameter $a$. The ellipsoidal metric
ansatz has also been used recently \cite{niko} for constructing the Kerr
solution in which ultimately the field equations are used.  Our focus here
is not to use the field equations but to deduce the metric by simply
appealing to the guiding principle of the incorporation of Newtonian
gravity and the velocity of light remaining constant in vacuum.

\section{Inclusion of charge: Kerr-Newman metric}

It is also possible to include electric charge on the black hole. For
this let us first consider the case of static black hole which would later be
generalized to rotating black hole. Note that our guiding principle
leads to the metric for static black hole as
$g_{tt} = (1+2\Phi) = -g_{rr}^{-1}$ where $\Phi$ is the Newtonian
potential which is $-M/r$ for the neutral source. This is the metric of the Schwarzschild
black hole. The question is what
should it be when charge $Q$ is put on it? The presence of charge produces
electric field energy spread over all space from the
source to infinity. Since gravity has universal linkage, electric field energy must also be included in gravitational field of charged source. If we wish to evaluate potential at some radius
$r$, the electric field energy lying outside the radius $r$ would be
$Q^2/2r$ which should be subtracted from the total mass $M$ of the
source \cite{d2}. This is because it is the energy interior to $r$ which
is only relevant for the potential and hence we write $\Phi = -(M-Q^2/2r)/r$. This is
then the Reissner-Nordstr\"{o}m metric of charged black hole. \\

Now let us similarly put charge on rotating black hole. For this
we need only to incorporate contribution of electric field energy to rotating black hole metric.
For this all the steps as above for rotating black hole remain true except for the
function $\psi(r)$ in $f(r) = g(r) = r^2+a^2+\psi$.  For that let us focus on the general potential term
for $\theta=0$
\begin{equation}
\frac{f(r)}{\rho^2} = \frac{r^2+a^2+\psi}{r^2+a^2} = 1 + \frac{\psi}{r^2+a^2} \approx 1 + \frac{\psi}{r^2} = 1 + 2\Phi.
\end{equation}
which means $\Phi = \psi/2r^2$. This should be the same as that for static charged black hole because the role of $a$ is
fully fixed by the ellipsoidal symmetrty and hence we wtite
\begin{equation}
\psi = 2r^2\Phi = - 2r^2(\frac{M - Q^2/2r}{r}) = -2Mr + Q^2.
\end{equation}
With this $f(r) = g(r) = r^2+a^2-2Mr+Q^2$ put in Eq. (5), we have obtained the Kerr-Newman metric
of charged rotating black hole. Note that role of rotation is entirely taken care of by
the ellipsoidal symmetry of space.

\section{Discussion}

The main purpose of this note is to demonstrate the fact that it is
possible to obtain the black hole metrics without solving the field
equations simply by appealing to a general guiding principle which points to
the covering properties of Einstein gravity over Newtonian gravity. The
symmetry part of the metric is dictated by the equipotential surfaces
which are spheres and ellipsoids for static and rotating black holes
respectively. The rest is quite uniquely fixed by incorporation of
velocity of light being constant and of Newtonian gravity in the first
approximation. This is a very novel and simple way of obtaining the
black hole metrics. \\

As pointed out by the author earlier \cite{d}, it is however remarkable that both the covering properties of
Einstein gravity are fully accounted for by curvature of space leaving however the
Newtonian inverse square law intact. The contribution of gravitational self energy is to curve
space\cite{d1} which is exactly what is required for light to feel gravity without having
to change its velocity. Thus gravitational field energy gravitates not as energy momentum distributuion
on the right of the Einstein equation but rather more subtly by curving the space. That is why
the gravitational force law remains firm and intact. Since the force law is the same, particle orbits basically retain their elliptic character. The curved nature of space can then only make orbits precess causing
perihelion shift. This is exactly what happens which has recently been demonstrated\cite{abra} beautifully by computing the Mercury perihelion advance in Newtonian gravity that requires space to be curved. It should also be noted that what is termed as bending of
light is actually a measure of bending of space for which light is simply used as a probe. This is because light cannot bend of its own as it experiences no acceleration like massive particles and hence
it truthfully follows the curved geometry of space thereby measuring bending of space. The point to note is what bends then is space not light \cite{d0}! \\

Further note that in the Einstein gravity, potential is exactly given by $\Phi = -M/r$ without the freedom of adding a constant to it. It can therefore vanish only at
infinity and nowhere else and so it is determined absolutely. As in Newtonian gravity we can't add a constant to it \cite{d1} because it produces non trivial gravitational dynamics representing energy momentum distribution due to a global monopole \cite{bv}. In contrast to the rest of physics, it is remarkable that constant potential in Einstein gravity is dynamically non-trivial. How do we
understand this absolute character of potential? Though black hole is an isolated
source but gravitational field it creates is spread all over the space and it also gravitates through space curvature. As a matter of fact empty space surrounding black hole is therefore not entirely empty gravitationally and it is reflected in space being curved instead of flat. Hence potential can vanish only when space turns flat and gravitational field goes to zero. That happens only at infinity and that is why potential vanishes at infinity and nowhere else. This is how it is determined absolutely\cite{d}. This is a very subtle and important conceptual point which is generally lost sight of or not adequately emphasized. It is also a distinguishing feature of the Einstein gravity. \\

It is the universal linkage of gravity to all particles including massless ones that dictates that spacetime can no longer remain an inert background but it has to imbibe the gravitational dynamics in its structure curved spacetime \cite{d0}. Thus passage from Newton to Einstein is a revolutionary enlargement of framework.  It marks a profound break from the rest of physics as no other field makes such a demand on spacetime  structure. For all other fields, spacetime acts as a fixed background totally unmindful of what happens in it and how physical laws are prescribed and their dynamics played out. While for gravity, it is the spacetime curvature that determines its dynamics. Unlike Newton Einstein had no freedom to prescribe the inverse square law, it has all to follow from the curvature of spacetime. It is the differential geometry that dictates that the Riemann curvature satisfies the Bianchi identity which leadas to the Einstein Equation. This happens not only for the Einstein theory but also for higher order Lovelock theory involving a polynomial in Riemann curvature \cite{d3}. Thus gravitational dynamics is kind of self determined.\\

I believe that this way of intuitive reasoning and formulating a general
guiding principle to deduce results in covering theory from the
existing one is not only simple and elegant but is very enlightening
and insightful. This is how one builds up intuition for the new theory
by understanding it as a logical and physical continuation of the earlier
one. Of course this may not always work in very complex and involved
situations. However I for one would never be fully happy and satisfied
unless I understand things at their roots from first principle.

%======================================%
%<<<<<<<<<<<<< REFERENCES >>>>>>>>>>>>>%
%======================================%

\end{document}